\documentclass{jfm}

\usepackage{graphicx}
\usepackage{epstopdf,epsfig}
\usepackage{newtxtext}
\usepackage{newtxmath}
\usepackage{natbib}
\usepackage{hyperref}
\hypersetup{
    colorlinks = true,
    urlcolor   = blue,
    citecolor  = black,
}

\usepackage{bm}
\usepackage{amsmath}

\usepackage{framed}
\usepackage{algorithm}
\usepackage{algorithmicx}
\usepackage{algcompatible}
\usepackage{lipsum}
\usepackage{bm}
\usepackage{xcolor}

\def\dashL{\bm{\mbox{--~--~--}}}

\def\Lbox{\mbox{---}~{\hspace*{-.1in} $\square$}\hspace*{-.1in}~\mbox{---}}
\def\Lcirc{\mbox{---}~{\hspace*{-.1in}$\circ$}\hspace*{-.1in}~\mbox{---}}
\def\Ltriangle{\mbox{---}~{\hspace*{-.1in}$\triangle$}\hspace*{-.1in}~\mbox{---}}

\def\Lcross{\mbox{---}~{\hspace*{-.14in} $\times$}\hspace*{-.1in}~\mbox{---}}

\newcommand{\RomanNumeralCaps}[1]
\nolinenumbers

\title{Role of triad interactions in spectral evolution of surface gravity waves in deep water}

\author{Zhou Zhang\aff{1}
  \and
  Yulin Pan\aff{1}
  \corresp{\email{yulinpan@umich.edu}}}

\affiliation{\aff{1}Department of Naval Architecture and Marine Engineering, University of Michigan, Ann Arbor, MI 48109, USA}

\begin{document}
\maketitle

\begin{abstract}
 It is generally accepted that the evolution of deep-water surface gravity wave spectrum is governed by quartet resonant and quasi-resonant interactions. However, it has also been reported in both experimental and computational studies that non-resonant triad interactions can play a role, e.g., generation of bound waves. In this study, we investigate the effects of triad and quartet interactions on the spectral evolution, by numerically tracking the contributions from quadratic and cubic terms in the dynamical equation. In a finite time interval, we find that the contribution from triad interactions follows the trend of that from quartet resonances (with comparable magnitude) for most wavenumbers, except that it peaks at low wavenumbers with very low initial energy. This result reveals two effects of triad interactions: (1) the non-resonant triad interactions can be connected to form quartet resonant interactions (hence exhibiting the comparable trend), which is a reflection of the normal form transformation applied in wave turbulence theory of surface gravity waves. (2) the triad interactions can fill energy into the low energy portion of the spectrum (low wavenumber part in this case) on a very fast time scale, with energy distributed in both bound and free modes at the same wavenumber. We further analyze the latter mechanism using a simple model with two initially active modes in the wavenumber domain. Analytical formulae are provided to describe the distribution of energy in free and bound modes with numerical validations.
\end{abstract}

\begin{keywords}
Authors should not enter keywords on the manuscript, as these must be chosen by the author during the online submission process and will then be added during the typesetting process (see \href{https://www.cambridge.org/core/journals/journal-of-fluid-mechanics/information/list-of-keywords}{Keyword PDF} for the full list).  Other classifications will be added at the same time.
\end{keywords}


\section{Introduction}\label{sec:intro}
Wave turbulence is a state of motion in systems characterized by nonlinear interactions among many waves at different scales. The long-term statistical properties of such systems are described theoretically in the framework of wave turbulence theory (WTT). In this framework, a kinetic equation (KE) can be derived to model the evolution of the wave spectrum under wave-wave interactions. Such interactions have to satisfy the resonance condition (or its approximation in the case of quasi-resonances):
\begin{equation}
    \bm{k}_1\pm\bm{k}_2\pm...\bm{k}_N=0,
\label{eq:resk}
\end{equation}
\begin{equation}
    \omega_1\pm\omega_2\pm...\omega_N=0,
\label{eq:resw}
\end{equation}
where $N$ is the number of modes in the interaction ($N=4$ for quartet and $N=3$ for triad), $\bm{k}_i$ is the wavenumber vector, which is related to frequency $\omega_i$ by the dispersion relation.

For surface gravity waves in deep water, it is widely accepted that the relevant interactions in the KE are quartet resonant interactions with $N=4$. This is associated with the fact that triad resonant interactions are prohibited by the dispersion relation $\omega \sim |\bm{k}|^{1/2}$. As a result, a normal form transformation (see detailed formulation in \cite{krasitskii1994reduced}) can be implemented to the dynamical equations to remove the quadratic-nonlinearity terms (corresponding to non-resonant triad interactions). The resulted equation, named Zakharov equation \citep{zakharov1968stability}, contains only the cubic terms, based on which the KE with $N=4$ can be further derived. This process implies that the effect of non-resonant triad interactions, instead of being null, can be merged into and understood at the level of quartet interactions (see other studies in the Fermi–Pasta–Ulam–Tsingou (FPUT) system \citep{ganapa2023quasiperiodicity}). This argument about triad interactions, however, remains on a theoretical level without any direct numerical demonstration.

\begin{figure}
  \centerline{\includegraphics[scale =0.6]{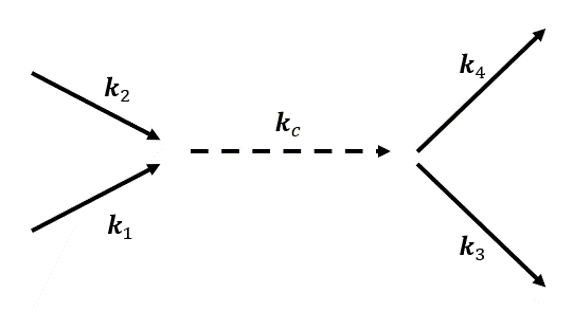}}
  \caption{A diagrammatic representation of a resonant quartet ($\bm{k}_1, \bm{k}_2, \bm{k}_3, \bm{k}_4$) formed by connection of two non-resonant triads satisfying $\bm{k}_1+\bm{k}_2=\bm{k}_c=\bm{k}_3+\bm{k}_4$, $\omega_1+\omega_2=\omega_c=\omega_3+\omega_4$.}
\label{fig:triadtoquartet}
\end{figure}

Another effect of triad interactions, which is discussed more often in literature, is the generation of bound modes. While such effect is absent from the Zakharov equation, it can indeed be expected from the original dynamical equation with quadratic nonlinearity. Given two free modes $\bm{k}_1$ and $\bm{k}_2$, the triad non-resonant interaction is supposed to generate another mode $\bm{k}_3$ with $\omega_3$ not satisfying the dispersion relation, hence named a bound mode. Such bound modes can be observed in the wavenumber-frequency spectrum as branches away from the dispersion relation curves, which are reported and discussed in many experimental and numerical studies \citep{herbert2010observation,cobelli2011different,aubourg2017three,campagne2019identifying,krogstad2010interpretations,taklo2015measurement,zhang2022numerical}. While the bound mode has been proposed as a building block connecting two triads to form a resonant quartet (see figure \ref{fig:triadtoquartet} and discussion in previous paragraph), a single triad is considered as purely non-resonant, i.e., it is only capable of generating bound modes, thus not contributing to the free-wave portion of the spectrum. 

Despite the above mentioned understanding on triad interactions, their effects on spectral evolution of gravity waves have not been completely understood, upon which questions often arise, say, in interpreting experimental data \cite[][]{aubourg2017three}. In this work, we aim to provide a comprehensive study on this problem. The core of this study is a numerical algorithm which we developed to directly track the contributions of quadratic and cubic terms in the dynamical equation (hence triad and quartet interactions) to spectral evolution. 

Through an analysis of the evolution of a typical gravity wave spectrum in a time interval, we find that the contributions from triad and quartet interactions share similar magnitude and trend for most wavenumbers, except that the former shows a much higher peak at small wavenumbers where the initial energy level is low. We point out that the comparable contributions from triad and quartet interactions at most wavenumbers is a direct manifestation of the normal form transformation, i.e., effect of triad interactions understood at the quartet level. We further study the rapid spectral evolution at low wavenumbers, which according to our results is due to non-resonant triad interactions. However, an analysis of the spectral content at low wavenumbers shows that there are significant energy in free waves in addition to bound waves, which seems to contradict with the previous understanding of non-resonant interactions (only generating bound modes). We reconcile this contradiction by providing a new understanding --- whenever a bound mode is generated, it is always accompanied by a free mode at the same wavenumber. This is in nature similar to the scenario of a frictionless pendulum forced at non-natural frequency, which leads to an oscillation at both forced and natural frequencies (the latter being a homogeneous solution). Based on this understanding, we analytically derive the energy ratio between bound and free modes in non-resonant triad interactions through a perturbation analysis. The obtained formula is finally validated favorably through simulations of a range of different triad interactions.

The paper is organized as follows. In \S \ref{sec:form}, we first review the problem formulation and normal form transformation. Then we present our algorithm to decompose quadratic and cubic terms in the gravity-wave dynamical equation, and the method to track their contributions in spectral evolution. In \S \ref{sec:results}, we analyze the evolution from a (tail-damped) JONSWAP spectrum, and elucidate the effect of triad interactions in this process, regarding the connection to form quartets and non-resonant interactions to distribute energy in generated bound and free modes. In \S \ref{sec:analysis}, we derive the analytical formula for the distribution of energy in triad interactions, with validations from direct simulations. Discussions and conclusions are provided in \S \ref{sec:conclu}.

\section{Formulation and Methodology}\label{sec:form}
We consider gravity waves on a two-dimensional (2D) free surface of an incompressible and irrotational ideal fluid with infinite depth. The flow field can be described by a velocity potential $\phi(\bm{x},z,t)$ satisfying Laplace's equation, where $\bm{x}=(x,y)$ is the horizontal coordinates, $z$ is the vertical coordinate and $t$ is time. The evolution of the surface elevation $\eta(\bm{x},t)$ and surface velocity potential $\psi(\bm{x},t)=\phi(\bm{x},z,t)|_{z=\eta}$ satisfy the Euler equations in Zakharov form \citep{zakharov1968stability}:
\begin{equation}
    \frac{\partial\eta}{\partial t}+\nabla_{\bm{x}}\eta\cdot\nabla_{\bm{x}}\psi-(1+\nabla_{\bm{x}}\eta\cdot\nabla_{\bm{x}}\eta)\phi_z=0
    \label{eq:eta}
\end{equation}
\begin{equation}
    \frac{\partial\psi}{\partial t}+\eta+\frac{1}{2}\nabla_{\bm{x}}\psi\cdot\nabla_{\bm{x}}\psi-\frac{1}{2}(1+\nabla_{\bm{x}}\eta\cdot\nabla_{\bm{x}}\eta)\phi_z^2=0
    \label{eq:psi}
\end{equation}
where $\phi_z(\bm{x},t)$ is the vertical velocity evaluated at the free surface, and $\nabla_{\bm{x}}=(\partial/\partial x,\partial/\partial y)$ denotes the horizontal gradient. With proper choice of mass and time units, the density and gravitational acceleration are both set to be unity so that they do not appear in \eqref{eq:psi}.

\subsection{Normal form transformation}
\label{sec:normalform}
Here we review the normal form transformation to remove the quadratic terms in \eqref{eq:eta} and \eqref{eq:psi}. The canonical variable for gravity waves is defined as 
\begin{equation}
    a_{\bm{k}}=\frac{1}{\sqrt{2}}(k^{-1/4} \hat{\eta}_{\bm{k}}+\mathrm{i}k^{1/4} \hat{\psi}_{\bm{k}}).
    \label{eq:ak}
\end{equation}
where a caret denotes the Fourier component in the wavenumber domain and $k=|\bm{k}|$. Writing \eqref{eq:eta} and \eqref{eq:psi} in terms of $a_{\bm{k}}$, we have (truncated up to cubic nonlinearity)
\begin{equation}
    \begin{aligned}
        \mathrm{i}\frac{\partial a_{\bm{k}}}{\partial t}=&\omega_{k}a_{\bm{k}} + \sum_{\bm{k}_1,\bm{k}_2} V^{[1]}_{012}a_1a_2\delta_{0-1-2} +\sum_{\bm{k}_1,\bm{k}_2} V^{[2]}_{012}a_1^*a_2\delta_{0+1-2} +\sum_{\bm{k}_1,\bm{k}_2} V^{[3]}_{012}a_1^*a_2^*\delta_{0+1+2} \\
        &+\sum_{\bm{k}_1,\bm{k}_2,\bm{k}_3} W^{[1]}_{0123}a_1a_2a_3\delta_{0-1-2-3} +\sum_{\bm{k}_1,\bm{k}_2,\bm{k}_3} W^{[2]}_{0123}a_1^*a_2a_3\delta_{0+1-2-3} \\
        &+\sum_{\bm{k}_1,\bm{k}_2,\bm{k}_3} W^{[3]}_{0123}a_1^*a_2^*a_3\delta_{0+1+2-3} +\sum_{\bm{k}_1,\bm{k}_2,\bm{k}_3} W^{[4]}_{0123}a_1^*a_2^*a_3^*\delta_{0+1+2+3},
    \end{aligned}
    \label{eq:dadt}
\end{equation}
where $V^{[n]}$, $W^{[n]}$ are interaction coefficients, and $\delta$ is the compact form of Kronecker delta function satisfying $\delta_{0\pm 1\pm 2}=\delta(\bm{k}\pm\bm{k}_1\pm\bm{k}_2)$ and $\delta_{0\pm 1\pm 2\pm 3}=\delta(\bm{k}\pm\bm{k}_1\pm\bm{k}_2\pm\bm{k}_3)$.

The goal is to introduce a near-identity transformation $a_{\bm{k}}\rightarrow b_{\bm{k}}$, in the form of 
\begin{equation}
    \begin{aligned}
        a_{\bm{k}}=&b_{\bm{k}}+\sum_{\bm{k}_1,\bm{k}_2} A^{[1]}_{012}b_1b_2\delta_{0-1-2} + \sum_{\bm{k}_1,\bm{k}_2} A^{[2]}_{012}b_1^*b_2\delta_{0+1-2} + \sum_{\bm{k}_1,\bm{k}_2} A^{[3]}_{012}b_1^*b_2^*\delta_{0+1+2} \\
        &+ \sum_{\bm{k}_1,\bm{k}_2,\bm{k}_3} B^{[1]}_{0123}b_1b_2b_3\delta_{0-1-2-3} + \sum_{\bm{k}_1,\bm{k}_2,\bm{k}_3} B^{[2]}_{0123}b_1^*b_2b_3\delta_{0+1-2-3} \\
        &+ \sum_{\bm{k}_1,\bm{k}_2,\bm{k}_3} B^{[3]}_{0123}b_1^*b_2^*b_3\delta_{0+1+2-3} + \sum_{\bm{k}_1,\bm{k}_2,\bm{k}_3} B^{[4]}_{0123}b_1^*b_2^*b_3^*\delta_{0+1+2+3}.
    \end{aligned}
    \label{eq:normalform}
\end{equation}
so that when we write the equation in terms of $b_{\bm{k}}$, it yields the form
\begin{equation}
    \mathrm{i}\frac{\partial b_{\bm{k}}}{\partial t} = \omega_kb_{\bm{k}} + \sum_{\bm{k}_1,\bm{k}_2,\bm{k}_3} T_{0123}b_1^*b_2b_3\delta_{0+1-2-3}.
    \label{eq:zakharoveq}
\end{equation}
Equation \eqref{eq:zakharoveq} is known as the Zakharov equation, based on which the KE associated only with quartet interactions can be derived. In order to obtain the Zakharov equation, the coefficients $A^{[n]}$ and $B^{[n]}$ in \eqref{eq:normalform} need to be chosen in a way such that the quadratic terms vanish under the transformation. The formulation of the coefficients can be found in \cite{krasitskii1994reduced}, with an example provided here:
\begin{equation}
    A^{[2]}_{012}= -\frac{V^{[2]}_{012}}{\omega_k+\omega_1-\omega_2},\ \text{for}\ \bm{k}_2=\bm{k}+\bm{k}_1,
    \label{eq:A2transform}
\end{equation}
where
\begin{equation}
    V^{[2]}_{012}=\frac{1}{2^{5/2}\pi}\left[\frac{k_2^{1/4}(\bm{k}\cdot\bm{k}_1+kk_1)}{(kk_1)^{1/4}} + \frac{k_1^{1/4}(\bm{k}\cdot\bm{k}_2-kk_2)}{(kk_2)^{1/4}} + \frac{k^{1/4}(\bm{k}_1\cdot\bm{k}_2-k_1k_2)}{(k_1k_2)^{1/4}} \right].
    \label{eq:v2coefficient}
\end{equation}
We remark that the denominator in \eqref{eq:A2transform} remains non-zero since there is no solution to \eqref{eq:resk} and \eqref{eq:resw} with $N=3$. This is the critical point allowing the normal form transformation to be implemented.

\subsection{Algorithm to decompose contributions from quadratic and cubic terms}\label{sec:decompose}
Our goal here is to compute numerically the contributions from quadratic and cubic terms in \eqref{eq:dadt} to the evolution of modal energy $e_{\bm{k}}=\omega_ka_{\bm{k}}a_{\bm{k}}^*$. However, a direct computation based on \eqref{eq:dadt} is very expensive with the convolutions, i.e., $O(N^2)$ and $O(N^3)$ computational complexity for quadratic and cubic terms with $N$ being the number of modes. To achieve an efficient $O(N\log N)$ computation, we need to go back to \eqref{eq:eta} and \eqref{eq:psi} in spectral domain where all terms can be computed making use of fast Fourier transform (FFT). For this purpose, we decompose \eqref{eq:eta} and \eqref{eq:psi} in spectral domain as
\begin{equation}
    \frac{\partial\hat{\eta}_{\bm{k}}}{\partial t}=\hat{L}_{\eta}+\hat{N}_{\eta}^{(2)}+\hat{N}_{\eta}^{(3)}+O(\epsilon^4),
    \label{eq:etadecompose}
\end{equation}
\begin{equation}
    \frac{\partial\hat{\psi}_{\bm{k}}}{\partial t}=\hat{L}_{\psi}+\hat{N}_{\psi}^{(2)}+\hat{N}_{\psi}^{(3)}+O(\epsilon^4),
    \label{eq:psidecompose}
\end{equation}
with linear terms 
\begin{equation}
    \hat{L}_{\eta}=\widehat{\phi_z^{(1)}},
    \label{eq:leta}
\end{equation}
\begin{equation}
    \hat{L}_{\psi}=-\hat{\eta}_{\bm{k}},
    \label{eq:lpsi}
\end{equation}
and nonlinear terms (up to cubic nonlinearity)
\begin{equation}
    \hat{N}_{\eta}^{(2)}=\widehat{\phi_z^{(2)}}-\widehat{(\nabla_{\bm{x}}\eta\cdot\nabla_{\bm{x}}\psi)},
    \label{eq:n2eta}
\end{equation}
\begin{equation}
    \hat{N}_{\psi}^{(2)}=\frac{1}{2}\widehat{[\phi_z^{(1)}]^2}-\frac{1}{2}\widehat{(\nabla_{\bm{x}}\psi\cdot\nabla_{\bm{x}}\psi)},
    \label{eq:n2psi}
\end{equation}
\begin{equation}
\hat{N}_{\eta}^{(3)}=\widehat{\phi_z^{(3)}}+\widehat{(\nabla_{\bm{x}}\eta\cdot\nabla_{\bm{x}}\eta\phi_z^{(1)})},
    \label{eq:n3eta}
\end{equation}
\begin{equation}
    \hat{N}_{\psi}^{(3)}=\widehat{\phi_z^{(1)}\phi_z^{(2)}}.
    \label{eq:n3psi}
\end{equation}   
In the above equations, $\phi_z^{(m)}$ represents $m$th-order-nonlinearity terms in $\phi_z$ when it is expressed as power series in $\eta$ and $\psi$, say, using the method in \cite{dommermuth1987high,pan2018high}. With such expressions of $\phi_z^{(m)}$, all terms in \eqref{eq:etadecompose} and \eqref{eq:psidecompose} can be evaluated straightforwardly employing FFT for the calculation of derivatives. 

We next connect \eqref{eq:etadecompose} and \eqref{eq:psidecompose} to the spectral energy evolution rate $r_{\bm{k}}=\partial e_{\bm{k}}/\partial t$, the quantity of our interest in the study. In terms of $\eta$ and $\psi$, $e_{\bm{k}}$ can be written as 
\begin{equation}
    e_{\bm{k}}=\frac{1}{2}(k\hat{\psi}_{\bm{k}}\hat{\psi}^*_{\bm{k}}+\hat{\eta}_{\bm{k}}\hat{\eta}^*_{\bm{k}}),
    \label{eq:ek}
\end{equation}
and $r_{\bm{k}}$ as 
\begin{equation}
    r_{\bm{k}}=\frac{1}{2}\left[k\left(\frac{\partial \hat{\psi}_{\bm{k}}}{\partial t}\hat{\psi}_{\bm{k}}^*+\frac{\partial \hat{\psi}_{\bm{k}}^*}{\partial t}\hat{\psi}_{\bm{k}}\right) + \left(\frac{\partial \hat{\eta}_{\bm{k}}}{\partial t}\hat{\eta}_{\bm{k}}^*+\frac{\partial \hat{\eta}_{\bm{k}}^*}{\partial t}\hat{\eta}_{\bm{k}}\right)\right],
    \label{eq:rk}
\end{equation}
(see similar implementation in Majda-Mclaughlin-Tabak model \citep{hrabski2022properties}) By substituting \eqref{eq:etadecompose} and \eqref{eq:psidecompose} to \eqref{eq:rk}, we obtain
\begin{equation}
    r^{(m)}_{\bm{k}}=\operatorname{Re}[k\hat{N}_{\psi}^{(m)}\hat{\psi}_{\bm{k}}^*+\hat{N}_{\eta}^{(m)}\hat{\eta}_{\bm{k}}^*],\quad m=2,3,...
    \label{eq:rkdecompose}
\end{equation}
where $r^{(m)}_{\bm{k}}$ gives the $k$-mode energy evolution rate due to $m$-th order nonlinearity in \eqref{eq:etadecompose} and \eqref{eq:psidecompose}. By definition, $r^{(1)}_{\bm{k}}=0$, while $r^{(2)}_{\bm{k}}$ and $r^{(3)}_{\bm{k}}$ correspond to the contributions from quadratic and cubic terms.

\subsection{Numerical setup}\label{sec:setup}
In this work, we simulate the evolution of a spectrum via \eqref{eq:eta} and \eqref{eq:psi}, with \eqref{eq:rkdecompose} implemented to track the contributions of triad and quartet interactions to spectral evolution. The initial condition for the simulation is chosen as a directional tail-damped JONSWAP spectrum \citep{hasselmann1973measurements}. Specifically, the spectrum in frequency-angle domain is written as $S(\omega,\theta)=G(\omega)D(\theta)$, where $\theta$ is the angle with respect to the positive $x$ direction. We set $G(\omega)=J(\omega)H(\omega)$. where $J(\omega)$ is a JONSWAP spectrum with the peak enhancement factor $\gamma=6$ (peak period $T_p$ and significant wave height $H_s$ specified later) and $H(\omega)$ is a tail-damped function in the form
\begin{equation}
    H(\omega)=
    \left\{
    \begin{array}{lc}
    1,& \omega^2< k_a\\
    e^{\lambda(\omega^2-k_a)},& \omega^2\geq k_a
    \end{array},
    \right.
\label{eq:damping}
\end{equation}
where $\lambda$ and $k_a$ are parameters controlling the damping rate and the effective range of damping, chosen as $k_a=40$ and $\lambda=-0.21$ in this study. The purpose of including $H(\omega)$ is to allow a larger room for evolution at the tail of the spectrum (since the JONSWAP spectrum is relatively close to stationary state at the tail). The angle spreading function $D(\theta)$ is chosen as a cosine-squared function in the form
\begin{equation}
    D(\theta)=
    \left\{
    \begin{array}{lc}
    \frac{2}{\pi}\cos^2\theta,& |\theta|\leq \pi/2\\
    0,& |\theta|>\pi/2
    \end{array}.
    \right.
\label{eq:direction}
\end{equation}
which satisfies $\int_{-\pi}^{\pi}D(\theta)d\theta=1$. The simulation of \eqref{eq:eta} and \eqref{eq:psi} is conducted in a doubly periodic domain of size $2\pi\times 2\pi$ (corresponding to a fundamental wavenumber $k_0=1$) with $512\times 512$ modes. An order-consistent high-order spectral (HOS) method \citep{west1987new} is used to compute the nonlinear terms up to cubic nonlinearity. For the time integration, we apply an integration factor scheme to solve for the linear terms analytically, and use a 4-th order Runge-Kutta method to integrate the nonlinear terms explicitly (a detailed description of these time-marching schemes can be found in \cite{pan2020high} in the context of capillary waves).

To stabilize the simulation, we add dissipation terms $D_{\eta}(k)=\gamma_k\eta$ and $D_{\psi}(k)=\gamma_k\psi$ to the right hand side of \eqref{eq:eta} and \eqref{eq:psi}, respectively, with the dissipation coefficients defined as $\gamma_k=\gamma_0(k/k_d)^{\nu}$. Such dissipation scheme has been applied in previous work \cite[][]{zhang2022numerical}. Here we use $\gamma_0=-50$, $k_d=115$ and $\nu=30$, and have tested that the resultant dissipation is negligible compared to the evolution due to nonlinear terms, so the evaluated $r^{(2)}_{\bm{k}}$ and $r^{(3)}_{\bm{k}}$ provide a correct view to the total evolution of the spectrum.

\section{Results}\label{sec:results}
We start by defining the angle-integrated modal energy and its evolution rate:
\begin{equation}
    E(k_r)=\sum_{|k-k_r|<\delta k}e_{\bm{k}},
    \label{eq:Ekr}
\end{equation}
\begin{equation}
    R^{(m)}(k_r)=\sum_{|k-k_r|<\delta k}r_{\bm{k}}^{(m)},
    \label{eq:Rkr}
\end{equation}
Where $\delta k$, chosen as 1, is a parameter characterizing the width of the (ring-shaped) region for the summation. Our interest is in the normalized and averaged energy evolution rate $|\overline{R^{(m)}}|/|\overline{E}|$ obtained from a time interval of spectral evolution, with the overbar denoting the time average in this interval. The denominator $|\overline{E}|$ accounts for the significant variation of energy level at different scales, which enables a fair comparison over all wavenumbers. We also remark that the quantity $|\overline{R^{(m)}}|/|\overline{E}|$ has a dimension of inverse time $[t]^{-1}$ and can be considered as an estimate of the reciprocal of nonlinear time scale.

\begin{figure}
  \centerline{\includegraphics[scale =0.48]{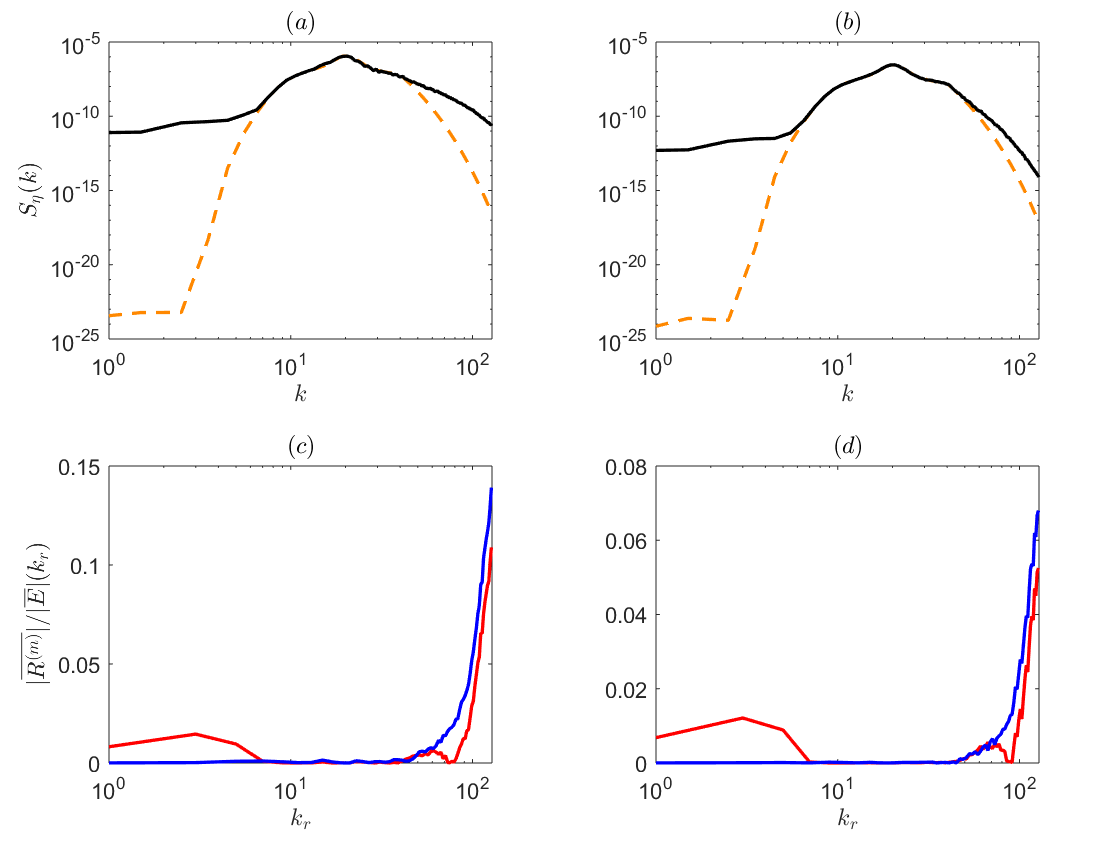}}
  \caption{Spectra of surface elevation obtained at $t=0$ ({\color{orange}\dashL}) and $t=80T_p$ ({\color{black}\rule[0.5ex]{0.5cm}{0.25pt}}) for (a) $\epsilon=0.1259$ and (b) $\epsilon=0.0629$; $|\overline{R^{(2)}}|/|\overline{E}|$ ({\color{red}\rule[0.5ex]{0.5cm}{0.25pt}}) and $|\overline{R^{(3)}}|/|\overline{E}|$ ({\color{blue}\rule[0.5ex]{0.5cm}{0.25pt}}) as functions of $k_r$ for (c) $\epsilon=0.1259$ and (d) $\epsilon=0.0629$.}
\label{fig:srek}
\end{figure}

\begin{figure}
  \centerline{\includegraphics[scale =0.48]{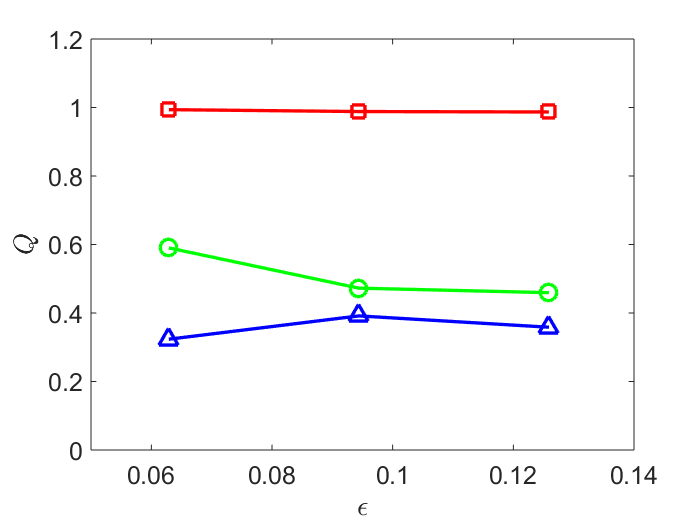}}
  \caption{Plots of $Q$ as functions of $\epsilon$ at $k_r=3$ ({\color{red}\Lbox}), $k_r=61$ ({\color{green}\Lcirc}) and $k_r=99$ ({\color{blue}\Ltriangle}).}
\label{fig:r2r3}
\end{figure}

Starting from the tail-damped JONSWAP spectrum introduced in \S \ref{sec:setup}, we perform simulations up to $80T_p$, where $T_p$ is the peak wave period corresponding to the
peak wavenumber $k_p=20$. The nonlinearity level $\epsilon=H_sk_p/2$ is controlled by varying $H_s$ in the initial spectrum. Figure \ref{fig:srek} shows the evolution of the surface-elevation spectrum $S_{\eta}(k)$ and the associated $|\overline{R^{(m)}}|/|\overline{E}|$ (with $m=2,3$) computed over $[0,80T_p]$ at two nonlinearity levels $\epsilon=0.1259$ and $0.0629$. From figure \ref{fig:srek}(a) and \ref{fig:srek}(b) we see significant spectral evolution occurring at small and large scales for both nonlinearity levels, with the evolution at higher nonlinearity more pronounced (a phenomenon related to the finite size effect which limits the spectral evolution at lower nonlinearity, see e.g. \cite{zhang2022numerical,pan2014direct}). The quantities $|\overline{R^{(m)}}|/|\overline{E}|$ shown in \ref{fig:srek}(c) and \ref{fig:srek}(d) reveal two observations which are true for both nonlinearities: (i) For most wavenumbers, especially closer to the tails of the spectrum, $|\overline{R^{(2)}}|/|\overline{E}|$ and $|\overline{R^{(3)}}|/|\overline{E}|$ show a comparable amplitude and similar trend. (ii) At small wavenumbers, $|\overline{R^{(2)}}|/|\overline{E}|$ is dominant to drive the spectral evolution. These two observations actually hold for more nonlinearity levels we have tested. To see this, we define another quantity 
\begin{equation}
    Q=\frac{|\overline{R^{(2)}}|}{|\overline{R^{(2)}}|+|\overline{R^{(3)}}|},
    \label{eq:Q23}
\end{equation}
to measure the relative intensity of triad interactions. Figure \ref{fig:r2r3} plots $Q(k_r)$ as a function of $\epsilon$ for $k_r=3$, $61$ and $99$. It is clear that both facts can be observed, in terms of $Q\approx 1$ for $k_r=3$ for observation (i) and $Q=0.3\sim 0.6$ for $k_r=61$ and $99$ for observation (ii). We next discuss in detail the mechanisms associated with the two facts.

\subsection{Discussion for observation (i)}
The observation that $|\overline{R^{(2)}}|\sim |\overline{R^{(3)}}|$ for most wavenumbers is in fact a manifestation of the normal form transformation discussed in \S \ref{sec:normalform}. In particular, for any resonant quartet, there are two ways for energy transfer to occur. The first way is obviously from cubic terms in the governing equations \eqref{eq:eta} and \eqref{eq:psi}. The second way is through quadratic terms, whose effect on spectral evolution is equivalent to the corresponding part in $T_{0123}$ in \eqref{eq:zakharoveq} obtained from the normal form transformation. As a result, the effects of both quadratic and cubic terms can be seen on the quartet level, leading to comparable trend and magnitude between $|\overline{R^{(2)}}|$ and $|\overline{R^{(3)}}|$ in most wavenumbers.

The effect of quadratic term on the quartet level can also be understood in a diagrammatic way. As shown in figure \ref{fig:triadtoquartet}, for any resonant quartet consisting of modes $\bm{k}_1$, $\bm{k}_2$, $\bm{k}_3$ and $\bm{k}_4$, one can always find a bridging mode $\bm{k}_c=\bm{k}_1+\bm{k}_2=\bm{k}_3+\bm{k}_4$, so that the resonant quartet is decomposed into two non-resonant triads $(\bm{k}_1,\bm{k}_2,\bm{k}_c)$ and $(\bm{k}_3,\bm{k}_4,\bm{k}_c)$. Moreover, the bridging mode is associated with the same bound frequency in the two triads, simply because $\omega_c=\omega_1+\omega_2=\omega_3+\omega_4$. This ensures that the (frequency) oscillation of the bridging mode generated from the first triad can be perfectly transferred to second triad, which is a critical condition for the transfer to be established. We note that this condition is only satisfied when $\bm{k}_1$, $\bm{k}_2$, $\bm{k}_3$ and $\bm{k}_4$ form a resonant quartet, which is the only situation (as opposed to non-resonant quartets) that such triad-induced transfer becomes effective.

Furthermore, as seen in figure \ref{fig:srek}, the relative contributions of $|\overline{R^{(2)}}|$ and $|\overline{R^{(3)}}|$ to the tail of the spectra remain similar at the two nonlinearity levels even though finite size effect limits the spectral evolution at lower nonlinearity. This suggests that both cubic- and quadratic-term-induced transfers on the quartet level are subject to the interplay between nonlinear broadening and wavenumber discreteness. When the nonlinearity level is sufficiently low in a finite domain, we can expect that quasi-resonances corresponding to both cubic and quadratic terms are depleted, leading to a wave field where the nonlinear interactions are dominated by exact resonances \citep{kartashova2008resonant,zhang2022forward}.


\subsection{Discussion for observation (ii)}
\label{sec:discussion2}
We now turn to the spectral evolution at small wavenumbers induced by triad interactions dominantly. Although the evolution seen in figure \ref{fig:srek}(a) and (b) look like an “inverse cascade”, the mechanism is completely different from the well-studied inverse cascade due to quartet resonant interactions, i.e., those resulting in the KZ solution \citep{annenkov2006direct,korotkevich2008simultaneous} or spectral peak downshift \citep{annenkov2006direct}. The time interval of our simulation, $80T_p$, is simply too short for these quartet-based processes to become relevant.

One may instead think that triad interactions resulting in the spectral evolution are quasi-resonant. Indeed, this is a question which often arises in the context of non-local triad interactions \citep{korotkevich2024non,onorato2009four}. In our case, a low-wavenumber mode $\bm{k}_L$ can be excited by two high-wavenumber modes $\bm{k}_1$ and $\bm{k}_2$ (say $\bm{k}_L+\bm{k}_1=\bm{k}_2$), so the interaction can be non-local. Now consider the limit $\bm{k}_L\rightarrow 0$, then the frequency mismatch $\Delta \omega \equiv \omega_L+\omega_1-\omega_2 \sim O(k_L^{1/2}) \rightarrow 0$. Will this trigger energy transfer by triad quasi-resonances and invalidate the normal form transformation due to small/zero divisor in \eqref{eq:A2transform}? It turns out that this is not a problem since as $k_L\rightarrow 0$, the numerator \eqref{eq:v2coefficient} of the transformation coefficient $V^{[2]}_{L12}\sim O(k_L^{3/4})\rightarrow 0$ faster than the denominator so the full term approaches zero instead of blowing up. Physically, this means that as $k_L\rightarrow 0$, the triad interaction coefficient approaches zero sufficiently fast so that there is no energy transfer by quasi-resonances. We note that the situation is different in shallow-water gravity waves as discussed in \cite{onorato2009four}. One can show that in the case of shallow-water waves, the denominator $\Delta \omega\sim O(k_L)$ and the numerator approaches zero following $O(k_L^{1/2})$, leading to a blow-up of the whole term. This means that for shallow water the normal form transformation should be applied with more caution, and the quasi-resonant triad interactions can indeed be relevant.

\begin{figure}
  \centerline{\includegraphics[scale =0.48]{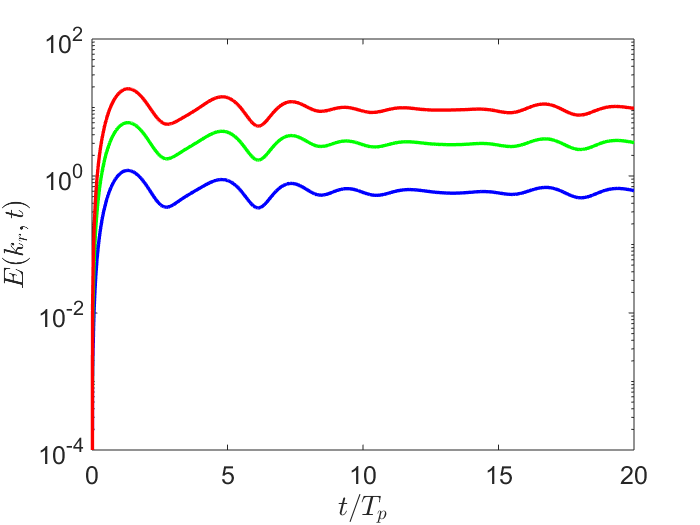}}
  \caption{Time evolution of $E(k_r,t)$ with $\epsilon=0.0629$ ({\color{blue}\rule[0.5ex]{0.5cm}{0.25pt}}), $\epsilon=0.0944$ ({\color{green}\rule[0.5ex]{0.5cm}{0.25pt}}) and $\epsilon=0.1259$ ({\color{red}\rule[0.5ex]{0.5cm}{0.25pt}}) at $k_r=3$.}
\label{fig:ekrt}
\end{figure}

\begin{figure}
  \centerline{\includegraphics[scale =0.48]{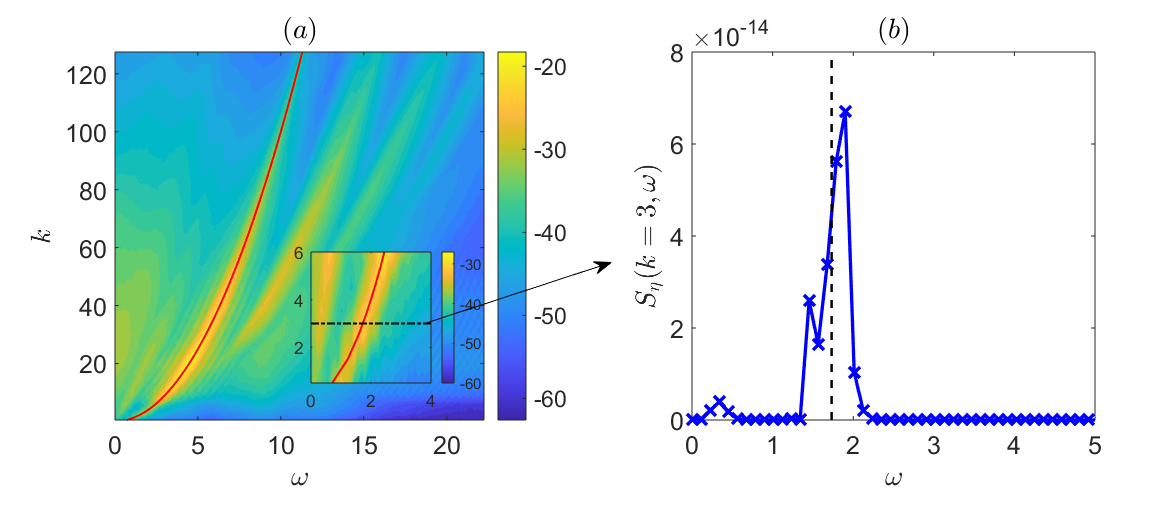}}
  \caption{(a) Wavenumber-frequency spectrum $S_\eta(k,\omega)$ in log scale with the dispersion relation marked by {\color{red}\rule[0.5ex]{0.5cm}{0.25pt}}, obtained with $\epsilon=0.0315$ and $T_L=40T_p$. A zoom-in view at small wavenumbers is shown to illustrate the portion of interest. (b) $S_\eta(k=3,\omega)$ as a function of $\omega$ ({\color{blue}\Lcross}),  with the corresponding free-mode frequency $\omega_f(k=3)$ marked by {\color{black}\dashL}.}
\label{fig:swkandsw}
\end{figure}

The absence of triad quasi-resonances is consistent with figure \ref{fig:ekrt} which plots the evolution of modal energy $E(k_r)$ with $k_r=3$ at three nonlinearity levels. We see that the modal energy rises from 0 to a stationary value in a very short linear time scale $O(T_p)$, which is a feature of non-resonant interactions instead of quasi-resonant interactions. Does this mean that the energy level at low wavenumbers is sustained by bound modes (according to the reasoning for non-resonant interactions)? To answer this question, we examine the wavenumber-frequency spectrum $S_{\eta}(k,\omega)$, defined as 
\begin{equation}
    S_\eta(k,\omega)=\int_0^{2\pi} |\hat{\eta}(\bm{k},\omega)|^2 k d\theta,
    \label{eq:Swk}
\end{equation}
where $\hat{\eta}(\bm{k},\omega)$ is the spatiotemporal Fourier transform of $\eta(\bm{x},t)$:
\begin{equation}
    \hat{\eta}(\bm{k},\omega)=\frac{1}{4\pi^2T_L}\iiint_{[0,T_L]\times[0,2\pi]\times[0,2\pi]} \eta(\bm{x},t)h_T(t)e^{-\mathrm{i}(\bm{k}\cdot\bm{x}-\omega t)}d\bm{x}dt,
\label{eq:etaKW}
\end{equation}
with $h_T(t)$ the Tukey window \citep{bloomfield2004fourier} of length $T_L$. Figure \ref{fig:swkandsw}(a) shows $S_{\eta}(k,\omega)$ with a zoom-in view of the low-wavenumber region. A more specific view is also provided in figure \ref{fig:swkandsw}(b) which plots $S_{\eta}(k=3,\omega)$ as a function of $\omega$ together with a vertical line denoting the free-mode frequency $\omega=\sqrt{3}$. Somewhat to our surprise, significant energy is located at the free mode, which is at least comparable to those away as bound modes. How do we reconcile the free mode generation and non-resonant interactions? In \S \ref{sec:analysis}, we will show explicitly that non-resonant interactions indeed generate both bound and free modes, with the distribution between these two components analytically solved with validations.

\section{Analytical study on triad non-resonant interactions with validations}
\label{sec:analysis}
We consider a simple model of a wave field consisting of two modes as the initial condition. In physical space, the initial field is described by
\begin{equation}
    \eta(\bm{x})=\eta_1(\bm{x})+\eta_2(\bm{x}),
    \label{eq:etainitial}
\end{equation}
\begin{equation}
    \psi(\bm{x})=\psi_1(\bm{x})+\psi_2(\bm{x}),
    \label{eq:psiinitial}
\end{equation}
with
\begin{equation}
    \eta_i(\bm{x})=\Tilde{A}_i\exp{(\mathrm{i}\bm{k}_i\cdot\bm{x})}+\Tilde{A}_i^*\exp{(-\mathrm{i}\bm{k}_i\cdot\bm{x})},
    \label{eq:eta2}
\end{equation}
\begin{equation}
    \psi_i(\bm{x})=-\frac{\mathrm{i}}{\omega_i} [\Tilde{A}_i\exp{(\mathrm{i}\bm{k}_i\cdot\bm{x})}-\Tilde{A}_i^*\exp{(-\mathrm{i}\bm{k}_i\cdot\bm{x})]},
    \label{eq:psi2}
\end{equation}
where $\Tilde{A}_i\in\mathbb{C}$, $\bm{k}_i=(k_{xi},k_{yi})\in \mathbb{R}^2$ and $\omega_i=k_i^{1/2}$ with $k_i=|\bm{k}_i|$ for $i=1,2$. This model allows two triad non-resonant interactions to occur, generating new modes at wavenumbers
\begin{equation}
    \bm{k}_{1+2}=\bm{k}_1+\bm{k}_2,
    \label{eq:kplus}
\end{equation}
\begin{equation}
    \bm{k}_{1-2}=\bm{k}_1-\bm{k}_2,
    \label{eq:kminus}
\end{equation}
respectively. Our goal is to show and elucidate the energy distribution between free and bound modes in this simpler problem. With the solution to this problem obtained, the spectral behavior at low wavenumbers observed in figures \ref{fig:ekrt} and \ref{fig:swkandsw} can be simply understood as the result of a collection of such non-resonant interactions. 

\begin{figure}
  \centerline{\includegraphics[scale =0.44]{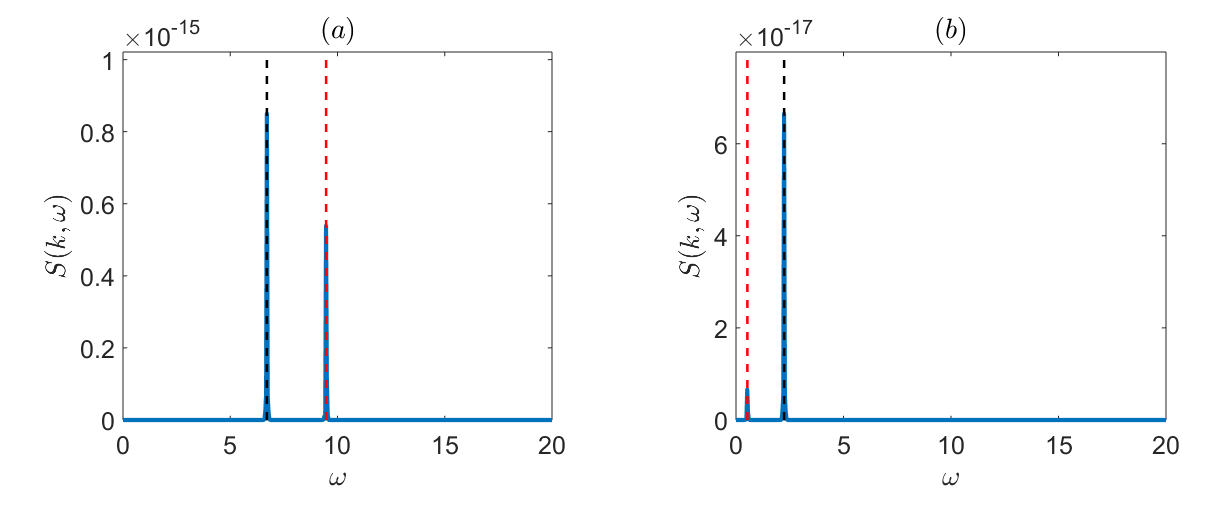}}
  \caption{$S_\eta(k,\omega)$ as functions of $\omega$ at (a) $k=k_{1+2}=45$ and (b) $k=k_{1-2}=5$ with $T_L=30T_0$. The corresponding free-mode frequencies $\omega_f$ are marked by {\color{black}\dashL} and bound-mode frequencies $\omega_b$ are marked by {\color{red}\dashL}.}
\label{fig:swk1k2}
\end{figure}

We first provide an illustrative HOS simulation by setting $\bm{k}_1=(25,0)$, $\bm{k}_2=(20,0)$, $|\Tilde{A}_1|=4\times 10^{-4}$, $|\Tilde{A}_2|=5\times 10^{-4}$ with initial phases $\arg{(\Tilde{A}_i)}$ assigned as random values within $[0,2\pi)$. The simulation is conducted up to quadratic nonlinearity to avoid potential higher-order effects. Our interest is in the wavenumbers $\bm{k}_{1+2}=(45,0)$ and $\bm{k}_{1-2}=(5,0)$. Figure \ref{fig:swk1k2} plots $S_\eta(k,\omega)$ (computed over $[0,30T_0]$ where $T_0=2\pi k_0^{-1/2}$ is the fundamental wave period) as a function of $\omega$ for (a) $k=k_{1+2}=45$ and (b) $k=k_{1-2}=5$. We see that for each wavenumber of interest, there are two peaks located at free-wave frequency ($\omega_f=\omega_{1+2}=6.71$ for $k_{1+2}$ and $\omega_f=\omega_{1-2}=2.24$ for $k_{1-2}$) and bound-wave frequency ($\omega_b=\omega_1+\omega_2=9.47$ for $k_{1+2}$ and $\omega_b=\omega_1-\omega_2=0.53$ for $k_{1-2}$). This is a clear evidence that non-resonant triad interactions can distribute comparable amount of energy between the free and bound modes. Defining $S_b=S_{\eta}(k_{1\pm 2},\omega_b)$ and $S_f=S_{\eta}(k_{1\pm 2},\omega_f)$, we can calculate that $S_b/S_f=0.63$ and $0.1$ for $k_{1+2}$ and $k_{1-2}$ in this case.

We next develop an analytical solution to describe quantitatively the distribution of energy, i.e., to obtain the value of $S_b/S_f$, for general configuration of $\bm{k}_1$ and $\bm{k}_2$. For this purpose, we perform a perturbation analysis based on \eqref{eq:eta} and \eqref{eq:psi}. As detailed in Appendix \ref{appA}, we set the first-order linear solution to match the linear wave field corresponding to the initial condition \eqref{eq:etainitial} and \eqref{eq:psiinitial}. Then we seek for second-order solution to describe the energy generation at free mode $S_f$ and bound mode $S_b$. A critical condition applied at the second order is a quiescent wave field for wavenumbers $\bm{k}_{1+2}$ and $\bm{k}_{1-2}$ at $t=0$, which allows the calculation of distribution of energy between the two frequencies at each wavenumber. The final solution yields the ratio between $S_b$ and $S_f$ at $\bm{k}_{1+2}$ and $\bm{k}_{1-2}$ respectively:
\begin{equation}
    (S_b/S_f)_{1+2}=\frac{|\Gamma_1+\Gamma_2|^2}{|\Gamma_1|^2+|\Gamma_2|^2},
    \label{eq:ratio1+2}
\end{equation}
\begin{equation}
    (S_b/S_f)_{1-2}=\frac{|\Lambda_1+\Lambda_2|^2}{|\Lambda_1|^2+|\Lambda_2|^2},
    \label{eq:ratio1-2}
\end{equation}
where 
\begin{equation}
    \Gamma_1=\frac{\alpha_{1,2}+\alpha_{2,1}+\beta_{1,2}+\beta_{2,1}}{\omega_1+\omega_2-\omega_{1+2}},
\end{equation}
\begin{equation}
    \Gamma_2=\frac{\alpha_{1,2}+\alpha_{2,1}-\beta_{1,2}-\beta_{2,1}}{\omega_1+\omega_2+\omega_{1+2}},
\end{equation}
\begin{equation}
    \Lambda_1=\frac{\alpha_{1,-2}-\alpha_{-2,1}-\beta_{1,-2}-\beta_{-2,1}}{\omega_1-\omega_2-\omega_{1-2}},
\end{equation}
\begin{equation}
    \Lambda_2=\frac{\alpha_{1,-2}-\alpha_{-2,1}+\beta_{1,-2}+\beta_{-2,1}}{\omega_1-\omega_2+\omega_{1-2}},
\end{equation}
with 
\begin{equation}
    \alpha_{i,j}=\frac{1}{\omega_i\omega_{i+j}}(\bm{k}_i\cdot\bm{k}_{i+j}-k_ik_{i+j}),
\end{equation}
\begin{equation}
    \beta_{i,j}=\frac{1}{2\omega_i\omega_j}(\bm{k}_i\cdot\bm{k}_j+k_ik_j).
\end{equation}
If we further consider the one-dimensional (1D) case where $\bm{k}_1=(k_{x1},0)$ and $\bm{k}_2=(k_{x2},0)$ are along the $x$ direction, then with some manipulations the expressions for $S_b/S_f$ are reduced to
\begin{equation}
    (S_b/S_f)_{1+2}=2\bigg/\left[\left(\frac{\omega_1+\omega_2}{\omega_{1+2}}\right)^{\mu}+1\right],
    \label{eq:ratio1d1+2}
\end{equation}
\begin{equation}
    (S_b/S_f)_{1-2}=2\bigg/\left[\left(\frac{\omega_{1-2}}{\omega_1-\omega_2}\right)^{\mu}+1\right],
    \label{eq:ratio1d1-2}
\end{equation}
where
\begin{equation}
    \mu=2\frac{\bm{k}_1\cdot\bm{k}_1}{k_1k_2}.
\end{equation}

Physical intuition on the coexistence of bound and free modes can also be established as a reflection of the analytical derivation: this phenomenon is physically similar to a friction-less pendulum subject to an external forcing at non-natural frequency. Both homogeneous solution at natural frequency (i.e. free mode) and inhomogeneous solution at forcing frequency (i.e., bound mode) can be excited, with the former persisting under the friction-less condition (which is usually not satisfied for vibration systems but indeed true for our wave problem at hand).

\begin{figure}
  \centerline{\includegraphics[scale =0.44]{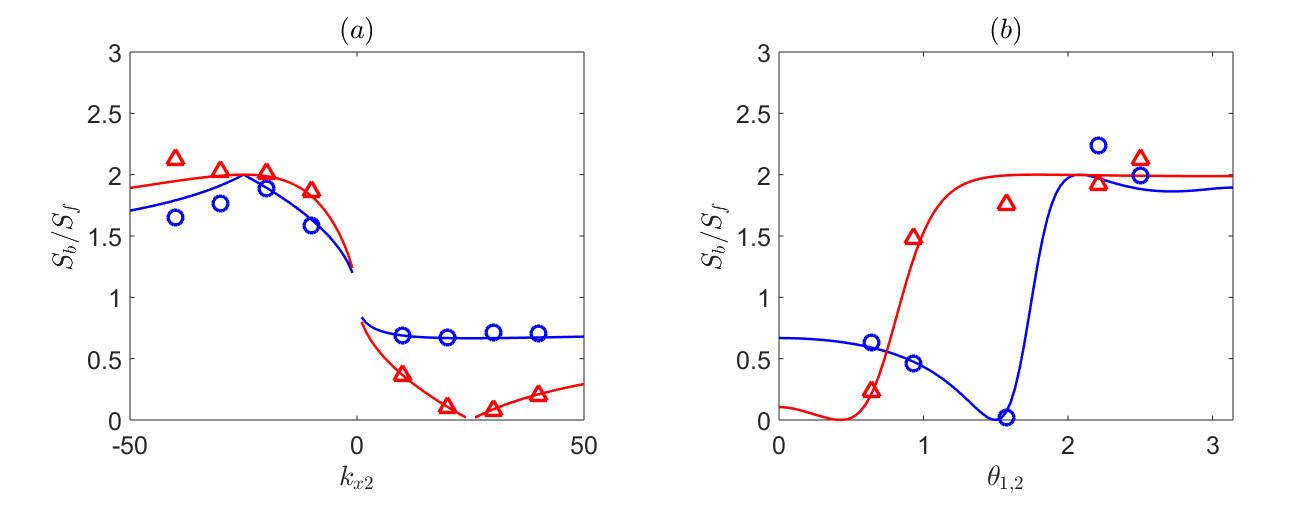}}
  \caption{Plots of $S_b/S_f$ as a function of (a) $k_{x2}$ in the 1D case with $\bm{k}_1=(25,0)$, $\bm{k}_2=(k_{x2},0)$ and $T_L=225T_0$ and (b) $\theta_{1,2}$ in the 2D case with $k_1=20$, $k_2=25$ and $T_L=20T_0$. For $k=k_{1+2}$, analytical solutions are shown by {\color{blue}\rule[0.5ex]{0.5cm}{0.25pt}} and numerical data are shown by {\color{blue}$\circ$}. For $k=k_{1-2}$, analytical solutions are shown by {\color{red}\rule[0.5ex]{0.5cm}{0.25pt}} and numerical data are shown by {\color{red}$\triangle$}.}
\label{fig:sbsf}
\end{figure}

We then conduct numerical validations for the analytical solution \eqref{eq:ratio1+2} and \eqref{eq:ratio1-2} via HOS. Two cases are considered: (1) a 1D case where we fix $\bm{k}_1=(25,0)$ and vary $\bm{k}_2=(k_{x2},0)$; (2) a 2D case where we fix $k_1=20$, $k_2=25$, and vary angle $\theta_{1,2}$ between them. The numerical solutions of $S_b/S_f$ are plotted in figure \ref{fig:sbsf} as a function of (a) $k_{x2}$ and (b) $\theta_{1,2}$ respectively, comparing against the analytical solutions \eqref{eq:ratio1+2} and \eqref{eq:ratio1-2} for 2D cases and \eqref{eq:ratio1d1+2} and \eqref{eq:ratio1d1-2} for 1D cases. We see overall a very good match. The minor deviation at a few points may be attributed to the finiteness of wave amplitude in the simulation and a finite time window in data processing which cannot exactly resolve $\omega_b$ and $\omega_f$. 

Finally, we notice that the analytical solution of $S_b/S_f$ is bounded in [0,2] in figure \ref{fig:sbsf}. This is in fact a general condition for $S_b/S_f$ that can be easily seen in the formulae \eqref{eq:ratio1+2} and \eqref{eq:ratio1-2} (simply because $0\leq (u+v)^2/(u^2+v^2)\leq 2$ for any $u,v\in \mathbb{R}$). This indicates that it is possible for the free-wave energy to be much larger than the bound-wave energy, but not vice versa. In other words, the free waves, instead of bound waves, should always be considered as a major generation from the non-resonant triad interactions. 

\section{Conclusions and Discussions}\label{sec:conclu}
In this paper, we present a study on the role of triad interactions in the spectral evolution of surface gravity waves in deep water. A decomposition technique is developed for Euler equations, which allows us to quantify the contributions from quadratic and cubic terms in spectral evolution. We find that, at most wavenumbers, the contribution from quadratic terms is comparable to that from cubic terms, reflecting the normal form transformation in theory to convert the quadratic to cubic terms. On the other hand, the quadratic terms dominates the spectral evolution at small wavenumbers with low initial energy levels. This is shown to be from non-resonant triad interactions which are found to distribute energy between free modes and bound modes. To understand this discovery of energy distribution, we start from a simple model with two existing modes and analytically solve for the generated modes. Our analytical solution to describe the distribution is validated by numerical simulations for a range of configurations, thus providing sound interpretation of our observation in the spectral evolution.

The results in this paper should not be considered as a challenge to the gravity-wave kinetic equation. The normal form transformation is indeed valid (see discussion for non-local triads in \S \ref{sec:discussion2}), although the rigorous mathematical justification only exists for 1D \citep{berti2023birkhoff}. However, some new understanding under the current framework should be noted. First, the contribution from triad interactions to spectral evolution is very significant, instead of null which may be expected by some researchers due to their non-resonant nature. Such contributions can come from resonant interactions connected by triads, which is comparable to cubic-term resonant interactions, or simply non-resonant triad interactions. In the latter case the non-resonant triad interactions can fill in the low-energy portion of the spectrum on a fast time scale, generating both free waves and bound waves. Only in a longer time scale, the spectral evolution becomes dominated by quartet interactions as described by the wave kinetic equation.

\backsection[Acknowledgements]{We thank Prof. Jalal Shatah, Prof. Miguel Onorato, Prof. Sergey Nazarenko, Prof. Fabio Pusateri and Dr. Michal Shavit for their valuable discussions and insights during the development of this work.}

\backsection[Funding]{We acknowledge the Simons Foundation for the funding support of this research.}

\backsection[Declaration of interests]{The authors report no conflict of interest.}


\backsection[Author ORCIDs]{Zhou Zhang, https://orcid.org/0000-0002-8820-4256; Yulin Pan, https://orcid.org/0000-0002-7504-8645}


\appendix

\section{Analytical solution of the ratio $S_b/S_f$ in \S \ref{sec:analysis}}
\label{appA}
To compute the ratio between the bound- and free-mode energy $S_b/S_f$ at a given wavenumber $\bm{k}$, we perform a perturbation analysis in this Appendix. This can be done by solving either the governing equations \eqref{eq:eta} and \eqref{eq:psi} directly or the equation for the complex wave amplitude $a_{\bm{k}}$ followed by the calculation of $\hat{\eta}_{\bm{k}}$. We have checked that both approaches are equivalent and give the same analytical result. Here, we follow the former way starting from the transformation of \eqref{eq:eta} and \eqref{eq:psi} into the wavenumber domain with a 2D spatial Fourier transform. The key procedure of this process is to express the surface vertical velocity $\phi_z$ in terms of $\eta$ and $\psi$ in the wavenumber domain, which is based on the small steepness assumption (i.e. $k\eta\sim O(\epsilon)\ll 1$). The details of this process can be found in \cite{mei2005theory,nazarenko2016wave} (also see \cite{pan2017understanding} in the context of capillary waves). For our purpose of investigating triad interactions, we consider the equations truncated up to $O(\epsilon^2)$:
\begin{equation}
    \frac{\partial \hat{\eta}_{\bm{k}}}{\partial t}=k\hat{\psi}_{\bm{k}}+\sum_{\bm{k}_1,\bm{k}_2} (\bm{k}_1\cdot\bm{k}-k_1k)\hat{\psi}_{\bm{k}_1}\hat{\eta}_{\bm{k}_2}\delta_{0-1-2},
    \label{eq:etakeq2}
\end{equation}
\begin{equation}
    \frac{\partial \hat{\psi}_{\bm{k}}}{\partial t}=-\hat{\eta}_{\bm{k}}+\frac{1}{2}\sum_{\bm{k}_1,\bm{k}_2} (\bm{k}_1\cdot\bm{k}_2+k_1k_2)\hat{\psi}_{\bm{k}_1}\hat{\psi}_{\bm{k}_2}\delta_{0-1-2},
    \label{eq:psikeq2}
\end{equation}
where $\bm{k}_1,\bm{k}_2\in\mathbb{Z}_L^2=2\pi\mathbb{Z}^2/L$ with $L$ being the size of the periodic square domain. We set the initial condition as in \eqref{eq:etainitial} to \eqref{eq:psi2}, expressed in wavenumber domain as
\begin{equation}
    \hat{\eta}_{\bm{k}}(t=0)=\Tilde{A}_1\delta_{0-1}+\Tilde{A}_2\delta_{0-2}+\Tilde{A}^*_1\delta_{0+1}+\Tilde{A}^*_2\delta_{0+2},
    \label{eq:etaic}
\end{equation}
\begin{equation}
    \hat{\psi}_{\bm{k}}(t=0)=-\frac{\mathrm{i}}{\omega_k}(\Tilde{A}_1\delta_{0-1}+\Tilde{A}_2\delta_{0-2}-\Tilde{A}^*_1\delta_{0+1}-\Tilde{A}^*_2\delta_{0+2}).
    \label{eq:psiic}
\end{equation}
We note that the Kronecker Delta functions on the right-hand side means that $\hat{\eta}_{\bm{k}}$ and $\hat{\psi}_{\bm{k}}$ take non-zero values only for $\bm{k}=\bm{k}_1,\bm{k}_2,-\bm{k}_1, -\bm{k}_2$, with the whole expression guaranteeing the two quantities to be real in physical domain. 

Our goal is to perform a perturbation analysis on \eqref{eq:etakeq2} and \eqref{eq:psikeq2}, with the first-order solution set as the wave field corresponding to initial conditions \eqref{eq:etaic} and \eqref{eq:psiic}. Then we seek second-order solution on wavenumber $\bm{k}_{1+2}$ and $\bm{k}_{1-2}$, which identifies $S_b/S_f$ as our quantity of interest. For this purpose, we first write the solutions of \eqref{eq:etakeq2} and \eqref{eq:psikeq2} as perturbation series in the wave steepness $\epsilon$:
\begin{equation}
    \hat{\eta}_{\bm{k}}=\hat{\eta}_{\bm{k}}^{(1)}+\hat{\eta}_{\bm{k}}^{(2)}+O(\epsilon^3),
    \label{eq:etaperturb}
\end{equation}
\begin{equation}
    \hat{\psi}_{\bm{k}}=\hat{\psi}_{\bm{k}}^{(1)}+\hat{\psi}_{\bm{k}}^{(2)}+O(\epsilon^3),
    \label{eq:psiperturb}
\end{equation}
where $\hat{\eta}_{\bm{k}}^{(m)},\hat{\psi}_{\bm{k}}^{(m)}\sim O(\epsilon^m)$ ($m=1,2,...$). By substituting \eqref{eq:etaperturb} and \eqref{eq:psiperturb} into \eqref{eq:etakeq2} and \eqref{eq:psikeq2}, we collect terms to obtain a set of equations at each order. The first-order linear equation reads:
\begin{equation}
    \frac{\partial \hat{\eta}_{\bm{k}}^{(1)}}{\partial t}=k\hat{\psi}^{(1)}_{\bm{k}},
    \label{eq:etalinear}
\end{equation}
\begin{equation}
    \frac{\partial \hat{\psi}_{\bm{k}}^{(1)}}{\partial t}=-\hat{\eta}^{(1)}_{\bm{k}}.
    \label{eq:psilinear}
\end{equation}
The solution to \eqref{eq:etalinear} and \eqref{eq:psilinear} are combination of linear waves propagating in the domain. Here we set this solution to be corresponding to the initial wave field \eqref{eq:etaic} and \eqref{eq:psiic}, written as
\begin{equation}
    \hat{\eta}_{\bm{k}}^{(1)}=(\Tilde{A}_1 \delta_{0-1} + \Tilde{A}_2 \delta_{0-2})e^{-\mathrm{i}\omega_kt} + (\Tilde{A}^*_1 \delta_{0+1} + \Tilde{A}^*_2 \delta_{0+2})e^{\mathrm{i}\omega_kt},
    \label{eq:etak1}
\end{equation}
\begin{equation}
    \hat{\psi}_{\bm{k}}^{(1)}=-\frac{\mathrm{i}}{\omega_k} [(\Tilde{A}_1 \delta_{0-1} + \Tilde{A}_2 \delta_{0-2})e^{-\mathrm{i}\omega_kt} - (\Tilde{A}^*_1 \delta_{0+1} + \Tilde{A}^*_2 \delta_{0+2})e^{\mathrm{i}\omega_kt}].
    \label{eq:psik1}
\end{equation}

Now we consider the second-order equations:
\begin{equation}
    \frac{\partial \hat{\eta}_{\bm{k}}^{(2)}}{\partial t}=k\hat{\psi}^{(2)}_{\bm{k}}+\sum_{\bm{k}_1,\bm{k}_2} (\bm{k}_1\cdot\bm{k}-k_1k)\hat{\psi}^{(1)}_{\bm{k}_1}\hat{\eta}^{(1)}_{\bm{k}_2}\delta_{0-1-2},
    \label{eq:etaquadratic}
\end{equation}
\begin{equation}
    \frac{\partial \hat{\psi}_{\bm{k}}^{(2)}}{\partial t}=-\hat{\eta}^{(2)}_{\bm{k}}+\frac{1}{2}\sum_{\bm{k}_1,\bm{k}_2} (\bm{k}_1\cdot\bm{k}_2+k_1k_2)\hat{\psi}^{(1)}_{\bm{k}_1}\hat{\psi}^{(1)}_{\bm{k}_2}\delta_{0-1-2}.
    \label{eq:psiquadratic}
\end{equation}
With the linear solutions $\hat{\eta}_{\bm{k}}^{(1)}$ and $\hat{\psi}_{\bm{k}}^{(1)}$ available, equations \eqref{eq:etaquadratic} and \eqref{eq:psiquadratic} are solved as a system of non-homogeneous linear differential equations. The corresponding second-order solutions at $\bm{k}=\bm{k}_{1+2}$ and $\bm{k}=\bm{k}_{1-2}$ can be expressed as
\begin{equation}
    \hat{\eta}_{\bm{k}}^{(2)}=\mathrm{i}\omega_k(C_1e^{-\mathrm{i}\omega_kt}-C_2e^{\mathrm{i}\omega_kt}+N_1-N_2),
    \label{eq:etak2}
\end{equation}
\begin{equation}
    \hat{\psi}_{\bm{k}}^{(2)}=C_1e^{-\mathrm{i}\omega_kt}+C_2e^{\mathrm{i}\omega_kt}+N_1+N_2,
    \label{eq:psik2}
\end{equation}
where $N_1$, $N_2$ have the following form:
\begin{equation}
    N_1=
    \left\{
    \begin{aligned}
        &\frac{1}{2} (-\alpha_{1,2}-\alpha_{2,1}-\beta_{1,2}-\beta_{2,1})\frac{\mathrm{i}\Tilde{A}_1\Tilde{A}_2}{\omega_1+\omega_2-\omega_k}e^{-\mathrm{i}(\omega_1+\omega_2)t},& \bm{k}=\bm{k}_{1+2}\\
        &\frac{1}{2} (-\alpha_{1,-2}-\alpha_{-2,1}+\beta_{1,-2}+\beta_{-2,1})\frac{\mathrm{i}\Tilde{A}_1\Tilde{A}^*_2}{\omega_1-\omega_2-\omega_k}e^{-\mathrm{i}(\omega_1-\omega_2)t},& \bm{k}=\bm{k}_{1-2}
    \end{aligned},
    \right.
    \label{eq:n1}
\end{equation}

\begin{equation}
    N_2=
    \left\{
    \begin{aligned}
        &\frac{1}{2} (\alpha_{1,2}+\alpha_{2,1}-\beta_{1,2}-\beta_{2,1})\frac{\mathrm{i}\Tilde{A}_1\Tilde{A}_2}{\omega_1+\omega_2+\omega_k}e^{-\mathrm{i}(\omega_1+\omega_2)t},& \bm{k}=\bm{k}_{1+2}\\
        &\frac{1}{2} (\alpha_{1,-2}+\alpha_{-2,1}+\beta_{1,-2}+\beta_{-2,1})\frac{\mathrm{i}\Tilde{A}_1\Tilde{A}^*_2}{\omega_1-\omega_2+\omega_k}e^{-\mathrm{i}(\omega_1-\omega_2)t},& \bm{k}=\bm{k}_{1-2}
    \end{aligned},
    \right.
    \label{eq:n2}
\end{equation}
with

\begin{equation}
    \alpha_{i,j}=\frac{1}{\omega_i\omega_{i+j}}(\bm{k}_i\cdot\bm{k}_{i+j}-k_ik_{i+j}),
    \label{eq:alphaij}
\end{equation}
\begin{equation}
    \beta_{i,j}=\frac{1}{2\omega_i\omega_j}(\bm{k}_i\cdot\bm{k}_j+k_ik_j).
    \label{eq:betaij}
\end{equation}

According to the initial conditions \eqref{eq:etaic} and \eqref{eq:psiic}, the wave field is quiescent except for modes $\bm{k}_1$ and $\bm{k}_2$. We therefore have the condition at $\bm{k}=\bm{k}_{1+2}$ and $\bm{k}=\bm{k}_{1-2}$:
\begin{equation}
    \hat{\eta}_{\bm{k}}^{(2)}(t=0)=0,
\end{equation}
\begin{equation}
    \hat{\psi}_{\bm{k}}^{(2)}(t=0)=0,
\end{equation}
from which the coefficients $C_1$ and $C_2$ in \eqref{eq:etak2} and \eqref{eq:psik2} are solved as
\begin{equation}
    C_1=-N_1(t=0),
    \label{eq:c1}
\end{equation}
\begin{equation}
    C_2=-N_2(t=0).
    \label{eq:c2}
\end{equation} 

With the second-order solutions obtained, we can now express $\hat{\eta}_{\bm{k}}$ at $\bm{k}_{1+2}$ and $\bm{k}_{1-2}$ in compact forms. For $\bm{k}_{1+2}$, we have
\begin{equation}
    \hat{\eta}_{1+2}=-\frac{\omega_{1+2}\Tilde{A}_1\Tilde{A}_2}{2}[\Gamma_1e^{-\mathrm{i}\omega_{1+2}t}+\Gamma_2e^{\mathrm{i}\omega_{1+2}t}-(\Gamma_1+\Gamma_2)e^{-\mathrm{i}(\omega_1+\omega_2)t}],
    \label{eq:eta1+2}
\end{equation}
where $\Gamma_i$ ($i=1,2$) are coefficients that can be evaluated explicitly from \eqref{eq:n1}, \eqref{eq:n2}, \eqref{eq:c1} and \eqref{eq:c2}. On the right hand side of \eqref{eq:eta1+2}, the first two terms correspond to free waves with frequency $\omega_{1+2}$ (propagating in opposite directions), while the third term corresponds to bound waves with frequency $\omega_1+\omega_2$. Therefore the ratio between energy in bound and free modes is expressed as
\begin{equation}
    (S_b/S_f)_{1+2}=\frac{|\Gamma_1+\Gamma_2|^2}{|\Gamma_1|^2+|\Gamma_2|^2},
    \label{eq:sbsf1+2}
\end{equation}
where
\begin{equation}
    \Gamma_1=\frac{\alpha_{1,2}+\alpha_{2,1}+\beta_{1,2}+\beta_{2,1}}{\omega_1+\omega_2-\omega_{1+2}},
\end{equation}
\begin{equation}
    \Gamma_2=\frac{\alpha_{1,2}+\alpha_{2,1}-\beta_{1,2}-\beta_{2,1}}{\omega_1+\omega_2+\omega_{1+2}}.
\end{equation}
For $\bm{k}_{1-2}$, similar procedures are applied yielding
\begin{equation}
    \hat{\eta}_{1-2}=-\frac{\omega_{1-2}\Tilde{A}_1\Tilde{A}^*_2}{2}[\Lambda_1e^{-\mathrm{i}\omega_{1-2}t}+\Lambda_2e^{\mathrm{i}\omega_{1-2}t}-(\Lambda_1+\Lambda_2)e^{-\mathrm{i}(\omega_1+\omega_2)t}],
    \label{eq:eta1-2}
\end{equation}
which gives the ratio
\begin{equation}
    (S_b/S_f)_{1-2}=\frac{|\Lambda_1+\Lambda_2|^2}{|\Lambda_1|^2+|\Lambda_2|^2},
    \label{eq:sbsf1-2}
\end{equation}
where
\begin{equation}
    \Lambda_1=\frac{\alpha_{1,-2}-\alpha_{-2,1}-\beta_{1,-2}-\beta_{-2,1}}{\omega_1-\omega_2-\omega_{1-2}},
\end{equation}
\begin{equation}
    \Lambda_2=\frac{\alpha_{1,-2}-\alpha_{-2,1}+\beta_{1,-2}+\beta_{-2,1}}{\omega_1-\omega_2+\omega_{1-2}}.
\end{equation}

Finally, we consider the results in the 1D case where $\bm{k}_1=(k_{x1},0)$ and $\bm{k}_2=(k_{x2},0)$ are along the $x$ direction. Then with some manipulations the expressions for $S_b/S_f$ are reduced to
\begin{equation}
    (S_b/S_f)_{1+2}=2\bigg/\left[\left(\frac{\omega_1+\omega_2}{\omega_{1+2}}\right)^{\mu}+1\right],
\end{equation}
\begin{equation}
    (S_b/S_f)_{1-2}=2\bigg/\left[\left(\frac{\omega_{1-2}}{\omega_1-\omega_2}\right)^{\mu}+1\right],
\end{equation}
where
\begin{equation}
    \mu=2\frac{\bm{k}_1\cdot\bm{k}_1}{k_1k_2}.
\end{equation}

\bibliographystyle{jfm}
\bibliography{jfm}

\end{document}